# Projection Super-resolution Based on Convolutional Neural Network for Computed Tomography

**Chao Tang, Wenkun Zhang, Ziheng Li, Ailong Cai, Linyuan Wang, Lei Li, Ningning Liang, Bin Yan\***

National Digital Switching System Engineering & Technological Research Centre, Zhengzhou, Henan, China, 450002

**Abstract**. The improvement of computed tomography (CT) image resolution is beneficial to the subsequent medical diagnosis, but it is usually limited by the scanning devices and great expense. Convolutional neural network (CNN)-based methods have achieved promising ability in super-resolution. However, existing methods mainly focus on the super-resolution of reconstructed image and do not fully explored the approach of super-resolution from projection-domain. In this paper, we studied the characteristic of projection and proposed a CNN-based super-resolution method to establish the mapping relationship of low- and high-resolution projection. The network label is high-resolution projection and the input is its corresponding interpolation data after down sampling. FDK algorithm is utilized for three-dimensional image reconstruction and one slice of reconstruction image is taken as an example to evaluate the performance of the proposed method. Qualitative and quantitative results show that the proposed method is potential to improve the resolution of projection and enables the reconstructed image with higher quality.

**\***Corresponding Author**,** E-mail: ybspace@hotmail.com

## 1 Introduction

Computed tomography (CT) technology has been widely applied to medical imaging, industrial non-destructive testing, and safety inspection because of its remarkable ability of visualizing the internal structure of scanned object[1]. In practice, high resolution imaging is necessary to enhance the fidelity of features of CT image. However, the image resolution of x-ray imaging system is constrained by the hardware device, e.g., x-ray focal spot size, detector element pitch, etc. Thus, how to get higher resolution images without changing the original hardware device has become a significant concern of researchers. In this paper, we proposed a super-resolution method to increase the x-ray imaging resolution by increasing the virtual sampling rate of detector element.

Super-resolution is a software technique of generating high-resolution images from low-resolution images[2]. Neural network-based super-resolution methods have outperformed traditional methods with its large dataset and high computing power graphics processing unit (GPU)[3]. As a pioneer work, Dong et al. proposed a super-resolution convolutional neural network (CNN) to learns an end-to-end mapping between low- and high-resolution images, with little extra pre/post-processing beyond the optimization[4]. In the field of CT imaging, several CNN-based methods are also proposed for image super-resolution since 2018. Park et al. proposed a deep CNN for brain CT image super-resolution[5]. Umehara et al. proposed the CNN-based method for enhancing image resolution in chest CT[6]. Hatvani et al. applied the CNN-based super-resolution to dental CT[7]. However, most of these methods use the strategy of image postprocessing to obtain high-resolution CT images. The information of projections has not received great attention in above studies and the study of projection super-resolution has not been fully explored.

In practice, some flat panel detectors such as Varian 2520D and 4030CB are equipped with two collection mode, i.e., 1×1 mode and 2×2 mode (www.varian.com), in most of cone-beam CT imaging system. The mode of 2×2 enables fast scanning and improves collection efficiency because this mode has a higher acquisition frame rate. However, the 2×2 mode is rarely adopted by the researcher because it obtains a low-resolution projection and the reconstructed image quality is poor. To solve this problem, this paper proposed a projection super-resolution method to establish the mapping relationship of low- and high-resolution projections based on the CNN architecture of Ref. 3. The workflow of projections super-resolution is shown in Figure 1. The existing high-resolution projections, i.e., ground truth, are used for the label of CNN. The low-resolution projections are obtained by down-sampling the ground truth projections and upscaling it via nearest interpolation. The input of CNN is the low-resolution projections. The higher resolution projections can be obtained through the trained CNN. This method will improve the virtual sampling rate of the detector element and enable the existing imaging system to achieve the image quality of 2×2 mode comparable to the 1×1 mode.

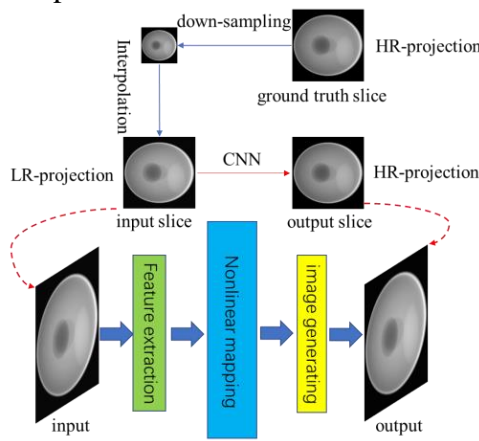

**Fig. 1** CNN architecture with data flow.

## 2  Method

The network architecture is shown in Fig. 2. The high-resolution projection is represented as $G$. The low-resolution projection is represented as $X$. The purpose of the network is to learn the mapping relationship $f$ from $X$ to $G$. The output $f(X)$ of network is as similar as possible to the ground truth high-resolution projection $G$. The proposed network is divided into three blocks: feature extraction, nonlinear mapping, and image generating. The 'Conv' layers of green represent convolution operation. The 'Relu' layers of yellow represent activation function.

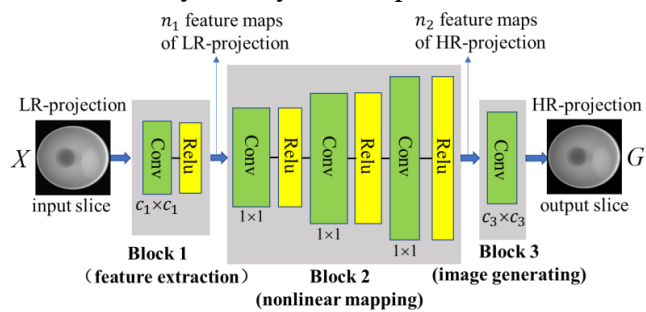

**Fig. 2** Network architecture of CNN. The 'Conv' layers of green represent convolution operation. The 'Relu' layers of yellow represent activation function.

The block of feature extraction is mainly a convolutional layer. This layer contains two parameters, i.e., $W_1$ is the weighting factor of the convolution kernel and $B_1$ is the bias. The first layer is represented by:

$$f_1(x) = \max(0, \; W_1 * X + B_1), \tag{1}$$

where * represents the convolution operation. The size of $W_1$ is $k \times c_1 \times c_1 \times n_1$, where $k$ is the number of channels in the image of input. The first layer performs $n_1$ convolution on the input, and the size of each convolution kernel is $c_1 \times c_1$. We can obtain the output of the first layer as $n_1$ feature maps. $B_1$ is an $n_1$-dimensional vector. We apply the rectified linear unit on the convolution kernel responses.

The block of nonlinear mapping mainly maps the $n_1$-dimensional vector obtained by the first block to the $n_2$-dimension. In order to increase the degree of nonlinear mapping as much as possible, we use three convolutional layers to gradually map $n_1$-dimensional vectors to $n_2$-dimensions, i.e., from $n_1$ to $n_i$, from $n_i$ to $n_j$, and finally from $n_j$ to $n_2$. The size of convolution kernel is $1 \times 1$. We represent each layer as follows:

$$f_i(x) = \max(0, \; W_i * f_1(x) + B_i), \tag{2}$$

$$f_j(x) = \max(0, \; W_j * f_i(x) + B_j), \tag{3}$$

$$f_2(x) = \max(0, \; W_2 * f_j(x) + B_2), \tag{4}$$

where the size of $W_i$ is $n_1 \times 1 \times 1 \times n_i$, the size of $W_j$ is $n_i \times 1 \times 1 \times n_j$, and the size of $W_2$ is $n_j \times 1 \times 1 \times n_2$. Correspondingly, the $B_i$ is an $n_i$-dimensional vector, the $B_j$ is an $n_j$-dimensional vector, and the $B_2$ is an $n_2$-dimensional vector.

Finally, the block of image generating handle with the $n_2$-dimensional feature vector obtained by the block of nonlinear mapping to obtain the output of network. The last layer is expressed as:

$$f(x) = W_3 * f_2(x) + B_3, \tag{5}$$

where the size of $W_3$ is $n_2 \times c_3 \times c_3 \times k$, and $B_3$ is a $k$-dimensional vector. Specifically, the parameters of the network are set as: $c_1 = 9$, $c_2 = 5$, $n_1 = 64$, $n_i = 48$, $n_j = 40$ and $n_2 = 32$.

Learning the mapping function $f$ required the estimation of network variables, which are weight and bias ($M$). The initial weight is a random value from a truncated normal distribution. The object of this study is that minimized loss function $L(v)$.

$$L(v) = \frac{1}{n} \sum_{i=0}^{n} \|f(X_i; v) - H_i\|^2, v = \{M_1, M_2 ... M_5\}, \tag{6}$$

where $n$ is the number of images in a mini-batch. The weight factor of each layer was updated using the error back-propagation with adaptive moment estimation optimizer[8], which is a stochastic optimization technique.

We simulate the three-dimensional shepp-logan phantom to verify the validity of our proposed method. The size of the phantom is set to 20mm × 20mm × 20mm. The distance from the X-ray source to the stage is set to 90mm, and the distance from the light source to the detector is set to 600mm. The size of the detector is set to 256mm × 256mm and the sampling width is 0.5mm. Then the resolution of the simulated virtual detector is calculated to be 3/40 mm. We set the scanning angle range from 0° to 360°, and collect a projection every 1°. In order to generate enough data, the ellipses inside the three-dimensional shepp-logan phantom is changed by a simple positional transformation, thereby generating different projection images. We obtained two groups of projection data of shepp-logan phantom for training and validating, each of which having 360 projection images. The size of the projection image is 512×512.

In the training phase, the ground truth projections $\{H_i\}$ are prepared as 33×33-pixel sub-images randomly cropped from the training data. To synthesize the low-resolution samples $\{X_i\}$, we blur a sub-image by a proper Gaussian kernel, down-sample it by the upscaling factor, and upscale it by the same factor via nearest interpolation. The 720 training projection images provide roughly 882000 sub-images. The sub-images are extracted from original images with stride 14. The training ($2\times10^6$ backpropagations) takes roughly 14 hours, on a Tesla k20x GPU.

## 3   Experimental Results

We used the projection data that is different from the training set to test the network. FDK algorithm[9] is used to generate three-dimensional CT images. The experimental results of the three-dimensional shepp-logan phantom are shown in Fig. 3. We compared the projection slices and reconstructed image slices separately.

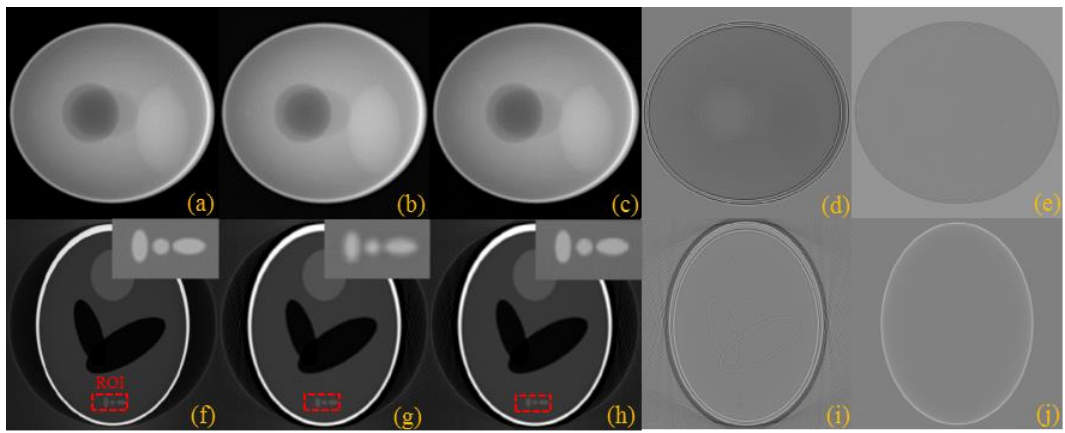

**Fig. 3** The super-resolution results of CT projection. (a)-(c) represents the projection of ground truth, bicubic interpolation, and network output, respectively. (d)-(e) are the projections errors of (b) and (c) with ground truth. (f)-(h) represents the reconstructed image of projection (a)-(c), respectively. (i)-(j) are the reconstruction errors of (g) and (h) with reconstruction image (f) that obtained from the ground truth projection. The display window of (a)-(c), (f)-(h) is [0, 10] and [0, 0.025], respectively. The display window of (d)-(e) and (i)-(j) is [-0.15, 0.1] and [-0.01, 0.01], respectively. The display window of ROIs represented by red rectangle in (f)-(h) is [0, 0.01].

We have adopted a four-fold scaling factor. From the experimental results, the proposed method will perform better on the contour. Peak signal to noise ratio (PSNR) and normalized root mean square error (NRMSE) are calculated to quantitatively evaluate the performance of proposed super-resolution method. The results of the numerical comparison are shown in Table 1. From the numerical results, our method is effective for super-resolution of CT image projections. The PSNR of projection is improved by nearly 28.2%, and the NRMSE is decreased by nearly 59.1%. The PSNR of ROI of image slice is improved by nearly 31.9%, and the NRMSE is decreased by nearly 55.1%.

Table 1 Evaluations of the results for super-resolution of CT image projections.

| Object | Scale | Bicubic | | CNN | |
|---|---|---|---|---|---|
| | | PSNR | NRMSE | PSNR | NRMSE |
| Projection | 4 | 27.5803 | 0.0418 | **35.3600** | **0.0171** |
| ROI of image slice | 4 | 21.8191 | 0.0811 | **28.7781** | **0.0364** |

## 4 Conclusion

This paper proposed a novel projection super-resolution method to realize high-resolution CT imaging. The proposed method can improve the virtual sampling rate of detector element and enable the existing imaging system to achieve the image quality of 2×2 mode comparable to the 1×1 mode. Experimental results verified that the performance of the proposed method in projection super-resolution. We will perform image super-resolution experiments with the actual CT data in future.

*Acknowledgments*

This work is supported by the National Natural Science Foundation of China (NSFC), Nos. 61601518.